\documentclass[aps, prfluids, showpacs, showkeys, notitlepage, amsmath, amssymb, floatfix, longbibliography, 12pt, tightenlines]{revtex4-2}

\usepackage{ulem}
\usepackage{svg}
\usepackage{float}
\usepackage{amsmath}
\usepackage{amssymb}
\usepackage[font=small, labelfont=bf, justification=justified]{caption}
\usepackage{subfig}
\usepackage{graphicx}
\usepackage{dcolumn}
\usepackage{bm}
\usepackage{hyperref}
\usepackage{booktabs}

\DeclareMathAlphabet{\altmathcal}{OMS}{cmsy}{m}{n}

\def\REF1#1{{ \textcolor{black}{#1}}}           
\def\REFEREE2#1{{\bfseries \textcolor{red}{#1}}}            

\begin{document}

\title{Behavior of hydrodynamic and magnetohydrodynamic turbulence in a rotating sphere with precession and dynamo action}
\author{M.~Etchevest}
\author{M.~Fontana}
\author{P.~Dmitruk}
\affiliation{Universidad de Buenos Aires, Facultad de Ciencias Exactas y Naturales, Departamento de F\'\i sica, \& IFIBA, CONICET, Ciudad Universitaria, Buenos Aires 1428, Argentina.}
\begin{abstract}
The effect of precession in a rotating sphere filled with fluid was studied with direct numerical simulations, both in the incompressible hydrodynamics (HD) and magnetohydrodynamics (MHD) scenarios. In both cases the asymptotic state and its dependence with both rotating and precession frequency was analyzed. \REF1{For the MHD case no self-sustaining dynamos were found  for the prograde precession case, whereas on the other hand a critical  retrograde precession frequency was found above which dynamo action is self-sustained.} It was also found that these correspond to small-scale dynamos with a developed turbulent regime. Furthermore, it is observed the presence of reversals of the magnetic dipole moment with greater waiting times between reversals for smaller precession frequencies.

\end{abstract}

\maketitle

\section{Introduction}

Different studies for the behavior of magnetic fields in stars and planets, and in particular the Earth, have supported the hypothesis that inside of these systems there is a flow of highly
conducting materials that produces magnetic induction \cite{Glatzmaiers1995,Braginsky1995,Roberts2000,Monchaux2009,Christensen2010}.
The usual framework for modelling this generation of magnetic fields from the kinetic energy of the flow
is that of magnetohydrodynamic (MHD) dynamos.
According to those studies, the rotatory movement of the planet can be one of the physical ingredients for
these chaotic, and often turbulent, flows \cite{stewartson1963,Busse1968,Noir2003,Tilgner2005}. 
Suggested drivers for these flows include thermal and/or chemical effects \REF1{and geometrical effects, like ellipticity, have been suggested to be relevant in the context of modelling the flows found in planet cores or stars \cite{Vidal2021}}. A distinctive forcing that was
also suggested is the precession effect in the otherwise constant 
rotation of the planet \cite{stewartson1963,Busse1968,Noir2003,Tilgner2005,Malkus1968,Keke}, a mechanism that can be responsible for the injection of kinetic energy
in the system. In the dynamics of the fluid (or magnetofluid) precession can be mathematically modelled through a term in the equations of motion (Navier-Stokes or MHD) that
is proportional to the time derivative of the angular velocity. 

\REF1{The precession effect in the full sphere was previously studied numerically, mostly in the laminar regime, as well as the transition to turbulence regime \cite{Lin2016}. Also, precessing spherical shells were analyzed in similar regimes \cite{Cebron2019}. Under these scenarios different types of dynamos were found, including stable dynamos, self-killing dynamos and intermittent dynamos. It was also suggested that the turbulent regime may lead to small-scale dynamo at the surface, both in the full sphere as in the spherical shell. Another subject that deserves some consideration when studying dynamo regimes is the choice of boundary conditions \cite{Fontana2022b}. In \cite{Lin2016} and \cite{Cebron2019} the poloidal component of the magnetic field smoothly matches with a potential field outside, whereas the assumption of an insulating boundary leads to the vanishing of the toroidal component at the borders of the domain. Adding the no-slip boundary condition to the aforementioned choice constitutes a geodynamo-like situation.}

For some dynamo systems that have a preferential direction, as is the case of rotating celestial bodies,
the magnetic fields arising from flow dynamics are known to undergo polarity reversals in their magnetic dipole
moment\cite{Babcock1961,Ponty2004,Olson2009,Amit2010,Monchaux2009}. We have previously studied these reversals in the context of
the magnetohydrodynamic equations with steady rotation \cite{Dmitruk2011,Dmitruk2014,Fontana2018}.
Here we consider the influence of a non-steady component in the angular velocity (i.e. of precession) on that phenomenon.

In this work we analyze the effect of precession on a conducting fluid inside a spherical cavity using 75 direct numerical simulations (DNS)
of the MHD equations. We first consider the purely hydrodynamic equations (i.e. no magnetic field) under rotation and precession at constant frequency,
and perform a parametric study to analyze the asymptotic behavior of the flow once a statistically stationary state is reached. 
Afterwards we proceed to analyze the  magnetohydrodynamic case, starting from an initial random  magnetic field and observing its evolution.
\REF1{We focus on the study of the turbulent regime and the presence of dynamo action, looking for self-sustaining dynamos, concentrating on the sensitivity to the rotation and precession frequency values. Furthermore, we analyze the existence of magnetic dipole reversals in the system and its behaviour with the presence of precession.}  

The organization of the paper is as follows. Section \ref{sec:level1} contains the model equations, including the initial conditions, and a description of the numerical method employed in the simulations. The results for these numerical simulations are found in section \ref{sec:Results}, distinguishing between the hydrodynamics (HD) and magnetohydrodynamics (MHD) cases. Finally, in the section \ref{conclusions} we summarize the main results of this study.

\section{\label{sec:level1}Simulation setup}
\subsection{\label{sec:level2}Model equations and numerical scheme}

 In this work we study a rotating spherical domain $V$ filled with an incompressible conducting fluid. The boundary is considered to be rotating in a non-steady way (i.e. precessing), with an angular velocity $\boldsymbol{\Omega} = \Omega_0[\cos(\gamma t)\sin(\alpha)\Hat{\boldsymbol x}+\sin(\gamma t)\sin(\alpha)\Hat{\boldsymbol y}+\cos(\alpha)\Hat{\boldsymbol z}]$, where $\Omega_0$ is the amplitude of the angular velocity, $\gamma$ its precession frequency and $\alpha$ its precession angle. The density of the fluid is taken to be uniform and equal to unity. In a non-inertial reference frame that rotates fixed with the domain's boundary, the usual magnetohydrodynamic equations can be written as follows
\begin{align}
    \label{NS}
    &\frac{\partial\boldsymbol{v}}{\partial t}=\boldsymbol{v}\times\boldsymbol{\omega}-\boldsymbol{\nabla}\mathcal{P}+\boldsymbol{J}\times\boldsymbol{B}+\nu\nabla^2\boldsymbol{v}-2\boldsymbol{\Omega}\times\boldsymbol{v}+\boldsymbol{r}\times\dot{\boldsymbol{\Omega}},\\
    \label{induction}
    &\frac{\partial\boldsymbol{B}}{\partial t}=\boldsymbol{\nabla}\times(\boldsymbol{v}\times\boldsymbol{B})+\eta\nabla^2\boldsymbol{B},\\
    \label{div_v}
    &\boldsymbol{\nabla}\cdot\boldsymbol{v}=0,\\
    \label{div_B}
    &\boldsymbol{\nabla}\cdot\boldsymbol{B}=0,
\end{align}
where $\boldsymbol{v}$, $\boldsymbol{\omega}$ and $\boldsymbol{B}$ are the velocity, vorticity and magnetic fields respectively. $\boldsymbol{J}=\boldsymbol{\nabla}\times\boldsymbol{B}$ is the current density, the total pressure is $\mathcal{P}$, and $\nu$ and $\eta$ are the kinematic viscosity and the magnetic diffusivity. ${\boldsymbol r}$ is the position vector, whereas ${\dot{\boldsymbol \Omega}}$ is the time derivative of the angular velocity. Note that the last term in the RHS of Eq. (\ref{NS}) correspond to the precession term. All quantities are expressed in dimensionless Alfvénic units.

The boundary conditions that we consider are that the normal components of $\boldsymbol{v}$, $\boldsymbol{B}$, $\boldsymbol{\omega}$ and $\boldsymbol{J}$ must all vanish in the surface of the spherical domain, $r=R=1$ (using the ratio of the sphere $R$ to normalize lengths). The vanishing normal component of $\boldsymbol{v}$ represents the fact that there is no mass flux across the surface. \REF1{Regarding electromagnetic boundary conditions, the vanishing normal components of $\boldsymbol{B}$ and $\boldsymbol{J}$ can be considered as modelling the situation where a thin layer of dielectric material ($\boldsymbol{J}\cdot \hat{\boldsymbol{r}}=0$) is coated on the outside by a perfect conductor ($\boldsymbol{B}\cdot \hat{\boldsymbol{r}}=0$). The condition that results in a vorticity field tangential to the surface is implied by, but does not imply, no-slip boundary conditions. A more extensive discussion on the choice of boundary conditions can be found in \cite{Mininni2006,Mininni2007}.
As mentioned before, different type of boundary conditions have been previously considered
in \cite{Lin2016,Cebron2019}. A related study in which we address the specific influence of the boundary conditions on the dynamo action has been considered in \cite{Fontana2022b} although in a different geometry.}

For the boundary conditions under consideration, the total energy balance is determined as 
\begin{equation}
    \frac{dE}{dt}= \int_V\boldsymbol{v}\cdot(\boldsymbol{ r}\times\dot{\boldsymbol \Omega})dV-2\nu Z-\int_V\eta|\boldsymbol{J}|^2dV+\int_S \nu(\boldsymbol{v}\times\boldsymbol{\omega})\cdot \hat{\boldsymbol n}dS ,
    \label{energy_balance}
\end{equation}
where $\hat{\boldsymbol n}$ is an outward-pointing unit vector normal to the sphere's surface. This equation can be obtained from the dynamical Eqs. (\ref{NS})-(\ref{div_B})
after multiplying by the velocity $\boldsymbol{v}$ and the magnetic field $\boldsymbol{B}$ respectively and integrating in volume space to obtain the total energy 
\begin{equation}
E = \frac{1}{2} \int_V \boldsymbol{v}^2 + \boldsymbol{B}^2 dV
\end{equation}
The right hand side of equation \ref{energy_balance} contains the injection energy $\epsilon$ given, by the precession term (first term containing the time derivative of the angular velocity), and dissipative terms like the fluid volume dissipation that involves the enstrophy $Z=1/2 \int_V|\boldsymbol{\omega}|^2dV$,
the magnetic field dissipation involving the current density. The surface term corresponds to the injection or dissipation of energy through the boundary. This energy balance equation will be the base for the
scaling models presented in the forthcoming sections. 

\REF1{Another important feature of magnetic fields is their topology. To study the symmetry degree of the magnetic field, the magnetic energy can be separated into each spherical harmonic contribution $E_l^B$,}
\REF1{
\begin{equation}
    E_l^B=\frac{1}{2}\sum_{q,m}|\xi_{qlm}^B|^2
\end{equation}}
\noindent\REF1{where the dipolar contribution to the magnetic energy is $E_1^B$, the quadrupolar one is $E_2^B$, and so on.
The magnetic dipole moment vector $\boldsymbol{m}$ is defined as}

\REF1{
\begin{equation}
    \boldsymbol{m}=\int_V \boldsymbol{r}\times\boldsymbol{J}dV,
\end{equation}}
\noindent\REF1{and the latitude angle of the dipole moment $\alpha$ is determined as}

\REF1{
\begin{equation}
    \alpha = \arctan \left(\frac{m_z}{\sqrt{m_x^2+m_y^2}}\right).
\end{equation}}

We numerically integrate Eqs. (\ref{NS})-(\ref{div_B}) for the magnetohydrodynamic case and with null $\boldsymbol{B}$ for the purely hydrodynamic case using the \texttt{SPHERE} code\cite{SPHERE}. In both scenarios we expand $\boldsymbol{v}$ and $\boldsymbol{B}$ in terms of Chandrasekhar-Kendall (CK) functions which constitute a spectral basis, as reported in \cite{Mininni2006,Mininni2007}. These functions are the eigenfunctions of the curl with linear eigenvalue, so they obey the following expression:
\begin{equation}
    \boldsymbol{\nabla}\times\boldsymbol{K}_i = k_i \boldsymbol{K}_i.
    \label{eigenf_curl}
\end{equation}
Equation (\ref{eigenf_curl}) can be converted into a vector Helmholtz equation, for which three indices \textit{q}, \textit{l} and \textit{m} are needed to express all the solutions. The vector fields $\boldsymbol{K}_i$ can be succinctly expressed as
\begin{equation}
    \boldsymbol{K}_{qlm}=k_{ql}(\boldsymbol{\nabla}\times\psi_{qlm}\hat{\boldsymbol{r}})+\boldsymbol{\nabla}\times(\boldsymbol{\nabla}\times\psi_{qlm}\hat{\boldsymbol{r}}),
\end{equation}
with $\psi_{qlm}$ a solution to the scalar Helmholtz equation. Considering that our spherical domain contains the origin, the collection $\psi_{qlm}$ is given by
\begin{equation}
    \psi_{qlm}=C_{ql}j_l(|k_{ql}|r)\mathrm{Y}_l^m(\theta,\varphi).
\end{equation}
where $j_l$ is the spherical Bessel function of order $l$ and $Y^m_l$ is the spherical harmonic of degree $l$ and order $m$. $C_{ql}$ is a normalization constant which we adjust for the base components to be orthonormal to each other with respect to the usual inner product. The indices are limited as follows: $l$ and $m$ must obey usual rules for spherical harmonic indexing, that is $l>1$ and $-l<m<l$, whereas the index $q$ can be any integer except for $q=0$. Considering this, if the fields are decomposed using the CK basis, their eigenvalues are $k_{ql}$ and satisfy $k_{-ql}=k_{ql}$ and hence CK functions with opposing values of the index $q$ correspond to fields with opposing helicity. Making an analogy with a Fourier decomposition, the eigenvalues $k_{ql}$ can be thought of as an analog of the wavenumber. The velocity and magnetic field can be therefore expressed as
\begin{align}
    \label{u_CK}
    \boldsymbol{v}(\boldsymbol{r},t)&=\sum_{\substack{q=-\infty\\q\ne0}}^\infty\sum_{l=1}^\infty\sum_{m=-l}^l\xi_{qlm}^v(t)\boldsymbol{K}_{qlm}(\boldsymbol{r}),\\
    \label{B_CK}
    \boldsymbol{B}(\boldsymbol{r},t)&=\sum_{\substack{q=-\infty\\q\ne0}}^\infty\sum_{l=1}^\infty\sum_{m=-l}^l\xi_{qlm}^B(t)\boldsymbol{K}_{qlm}(\boldsymbol{r}).
\end{align}
The coefficients $\xi^v_i$ and $\xi^B_i$ only depend on time and to obtain its evolution we use Eqs. (\ref{NS}) and (\ref{induction}) but with the fields expanded in the CK basis, obtaining the following set of ordinary differential equations
\begin{align}
    \label{vcoeff_evolution}
    \frac{d\xi_n^v}{dt}&=\sum_{ij}k_jI_{ij}^n\left(\xi_i^v\xi_j^v-\xi_i^B\xi_j^B \right)+2\sum_i\xi_i^v\boldsymbol{\Omega}\cdot\boldsymbol{\mathcal{O}}_i^n-\nu k_n^2\xi_n^v+\mathcal{B}_n \delta_{l,1},\\
    \label{Bcoeff_evolution}
    \frac{d\xi_n^B}{dt}&=\sum_{ij}k_nI_{ij}^n\xi_i^v\xi_j^B-\eta k_n^2\xi_n^B.
\end{align}
Here, for notation clarity, $n$, $i$ and $j$ each represent a $(q,l,m)$, and $I_{ij}^n$ and $\boldsymbol{\mathcal{O}}_i^n$ are coupling arrays which obey the following expressions
\begin{align}
    \label{I_ij^n}
    I_{ij}^n &= \int_V\boldsymbol{K}_n^*\cdot(\boldsymbol{K}_i\times\boldsymbol{K}_j)dV,\\
    \label{O_i^n}
    \boldsymbol{\mathcal{O}}_i^n&=\int_V\boldsymbol{K}_n^*\times\boldsymbol{K}_idV.
\end{align}
On the other hand, the term that comes from the forcing given by the precession is defined as
\begin{equation}
    \mathcal{B}_n=4\sqrt{\frac{\pi}{3}}C_{q,l}\operatorname{sg}(k_{q,l})j_l'(|k_{q,l}|)\left(\dot{\Omega}_z\delta_{m,0}-\frac{1}{\sqrt{2}}(\dot{\Omega}_x-i\dot{\Omega}_y)\delta_{m,1}\right)
    \label{Poincare_force_CK}
\end{equation}
with $\operatorname{sg}$ the sign function and $\delta$ the Kronecker delta.

To solve Eqs. (\ref{vcoeff_evolution}) and (\ref{Bcoeff_evolution}) in a computer, a numerical resolution $q_\text{max}$ and $l_\text{max}$ has to be chosen. Given a fixed resolution, all the  normalization constant $C_{ql}$, the eigenvalues $k_i$ and the coupling arrays $I_{ij}^n$ and $\boldsymbol{\mathcal{O}}_i^n$ are computed. These tables can then be stored and used for the computation of the time-dependent coefficients. Due to the high precision used in the spatial discretization (i.e. the CK basis) Eqs. (\ref{vcoeff_evolution}) and (\ref{Bcoeff_evolution}) are integrated in time employing a fourth-order Runge-Kutta scheme.

\subsection{\label{subsec:Simulations}Simulations performed}
We carried out 75 direct numerical simulations, varying different parameters, as explained below. All of the simulations used $q_\text{max}=l_\text{max}=7$, which implies that 882 expansion coefficients are evolved in time. The temporal resolution was $\Delta t=10^{-3}$. Half of the simulations were for the hydrodynamic case and the other half for the magnetohydrodynamic case.

The initial conditions for the HD case consisted of exciting only the following helical modes: 
\begin{equation}
    \xi^v_{qlm}\Big|_{t=0} = 0.3 \  \ \text{with}\
    \begin{cases}
    q = \pm 1, \pm 2\\
    l = 1,2 \\
    m = 1,2
    \end{cases}
    \label{initial_v}
\end{equation}
Here, the $0.3$ value was chosen \textit{ad hoc} seeking to approximately normalize the initial energy.

The angular speed $\Omega_0$, the precession frequency $\gamma$ and the kinematic viscosity $\nu$ were varied in each simulation. Table \ref{table_HD} contains the simulations that we show in the figures in Section \ref{sec:Results} with their corresponding names (ID) and relevant adimensional parameters. The different $\nu$ values considered were the following ones: $\nu = 0.01$, $\nu = 0.03$, $\nu = 0.06$ and finally $\nu = 0.1$. It should be noted that not all the simulations are contained in the table, for better understanding only those shown in the figures are included. Other parameters that appear in the table are the initial Reynolds number ($Re_0$), the final Reynolds number ($Re_f$), the Ekman number ($Ek$), the final Rossby number ($Ro_f$) and the Poincaré number ($Po$). All of them were calculated considering the sphere radius as the length scale. Those characteristic numbers differ according to the moment of the run in which the typical velocity is computed. The final velocity was determined as the time average of the flow speed in the time window where the kinetic energy is steady. The parameters can be calculated as:
\begin{equation} \label{adim_param}
\begin{aligned}
Re_f &= \frac{v_f R}{\nu}, \\
Ek &= \frac{\nu}{R^2 \Omega_0}, \\
Ro_f &= \frac{v_f}{R \Omega_0}, \\
Po &= \frac{|\gamma|}{\Omega_0}.
\end{aligned}
\end{equation}
The table also includes $Re_0$, which is defined similarly to  $Re_f$ but changing the final velocity $v_f$ for the initial velocity $v_0$. As indicated $Po$ is the ratio between the precession frequency and the angular velocity.

\begin{table}[H]
\begin{center}
\begin{tabular*}{.9\textwidth}{ c @{\extracolsep{\fill}} c c c c c c c }
\toprule
\toprule
ID & $\Omega_0$ & $\gamma$ & $Po$ & $Re_0$ & $Re_f$ & $Ek$ & $Ro_f$\\ \hline
\midrule
HD01 & $1$ & $0.1$ & $1.0\times10^{-1}$ & $1.0\times10^{2}$ & $1.2\times10^{0}$ & $1.0\times10^{-2}$ & $1.2\times10^{-2}$ \\
HD02 & $4$ & $0.1$ & $2.5\times10^{-2}$& $1.0\times10^{2}$ & $2.0\times10^{0}$ & $2.5\times10^{-3}$ & $5.0\times10^{-3}$ \\
HD03 & $10$ & $0.1$ & $1.0\times10^{-2}$ & $1.0\times10^{2}$ & $2.1\times10^{0}$ & $1.0\times10^{-3}$ & $2.1\times10^{-3}$\\
HD04 & $1$ & $1$ & $1.0\times10^{0}$ & $1.0\times10^{2}$ & $9.7\times10^{0}$ & $1.0\times10^{-2}$ & $9.7\times10^{-2}$\\
HD05 & $4$ & $1$ & $2.5\times10^{-1}$ & $1.0\times10^{2}$ & $1.7\times10^{1}$ & $2.5\times10^{-3}$ & $4.2\times10^{-2}$\\
HD06 & $10$ & $1$ &  $1.0\times10^{-1}$  & $1.0\times10^{2}$ & $2.0\times10^{1}$ & $1.0\times10^{-3}$ & $2.0\times10^{-2}$\\
HD07 & $1$ & $10$ & $1.0\times10^{1}$ & $1.0\times10^{2}$ & $1.8\times10^{1}$ & $1.0\times10^{-2}$  & $1.8\times10^{-1}$\\
HD08 & $4$ & $10$ & $2.5\times10^{0}$ & $1.0\times10^{2}$ & $5.7\times10^{1}$ & $2.5\times10^{-3}$ & $1.4\times10^{-1}$\\
HD09 & $10$ & $10$ & $1.0\times10^{0}$ & $1.0\times10^{2}$ & $1.1\times10^{2}$ & $1.0\times10^{-3}$ & $1.1\times10^{-1}$\\

HD10 & $1$ & $0.1$ & $1.0\times10^{-1}$ & $3.5\times10^{1}$ &  $4.1\times10^{-1}$ & $3.0\times10^{-2}$ & $1.2\times10^{-2}$ \\
HD11 & $4$ & $0.1$ & $2.5\times10^{-2}$ & $3.5\times10^{1}$ & $6.2\times10^{-1}$ & $7.5\times10^{-3}$ & $4.6\times10^{-3}$\\
HD12 & $10$ & $0.1$ & $1.0\times10^{-2}$ & $3.5\times10^{1}$ & $6.7\times10^{-1}$ & $3.0\times10^{-3}$ & $2.0\times10^{-3}$\\
HD13 & $1$ & $1$ &  $1.0\times10^{0}$ & $3.5\times10^{1}$ & $2.8\times10^{0}$ & $3.0\times10^{-2}$ & $8.4\times10^{-2}$\\
HD14 & $4$ & $1$ & $2.5\times10^{-1}$ & $3.5\times10^{1}$ & $5.2\times10^{0}$ & $7.5\times10^{-3}$ & $3.9\times10^{-2}$\\
HD15 & $10$ & $1$ & $1.0\times10^{-1}$ & $3.5\times10^{1}$ & $6.2\times10^{0}$ & $3.0\times10^{-3}$ & $1.9\times10^{-2}$\\
HD16 & $1$ & $10$ & $1.0\times10^{1}$ & $3.5\times10^{1}$ & $5.8\times10^{0}$ & $3.0\times10^{-2}$ & $1.7\times10^{-1}$\\
HD17 & $4$ & $10$ & $2.5\times10^{0}$ & $3.5\times10^{1}$ & $1.9\times10^{1}$ & $7.5\times10^{-3}$ & $1.4\times10^{-1}$\\
HD18 & $10$ & $10$ & $1.0\times10^{0}$ &  $3.5\times10^{1}$ & $3.5\times10^{1}$ & $3.0\times10^{-3}$ & $1.0\times10^{-1}$\\
\bottomrule
\bottomrule
\end{tabular*}
\caption{Runs corresponding to the HD case with their respective names (ID) and relevant adimensional parameters. The difference between the two Reynolds numbers is the time in which each one was calculated, the initial $Re_0$ and the final $Re_f$. The characteristic length considered to calculate all adimensional parameters was the sphere radius, and the initial velocity is of the order of unity.}
\label{table_HD}
\end{center}
\end{table}

On the other hand the MHD simulations were initially excited in the following modes for the velocity field
\begin{equation}
    \xi^v_{qlm}\Big|_{t=0} = 0.5 \  \ \text{with $q=3$}\
    \begin{cases}
    \text{if}\ l=1,2 \Rightarrow m=0,1 \\
    \text{if}\ l=3 \Rightarrow  m\in [0,3] 
    \end{cases}
    \label{initial_v_MHD}
\end{equation}
and for the magnetic field:
\begin{equation} \label{initial_B}
\begin{aligned}
\xi_{110}^B\Big|_{t=0}&=0.5, \\
\xi_{111}^B\Big|_{t=0}&=0.5 (1-i),
\end{aligned}
\end{equation}
in order to have fields with net initial helicity, a feature known to favor dynamo action. The corresponding Table \ref{table_MHD} shows the different runs for the MHD case. \REF1{A unit magnetic Prandtl number $P_m$ is prescribed, that is, the magnetic diffusivity $\eta$ is equal to the kinematic viscosity $\nu$ in all the simulations.
This differs from the typical values found in astrophysical scenarios, which are usually considered to be in the range $10^{-6} -- 10^{-3}$. However, based on recent studies \cite{Fontana2022b}, we expect that similar regimes to the ones we report here might be attainable provided the magnetic Reynolds number $Re_m$ is high enough.}
In contrast to the HD case, only the values of $\Omega_0$ and $\gamma$ were varied. In this case we also considered negative values of the precession frequency, an important fact that we discuss in Section \ref{sec:Results}.

\begin{table}[H]
\begin{center}
\begin{tabular*}{.9\textwidth}{ c @{\extracolsep{\fill}} c c c c c c c c }
\toprule
\toprule
ID & $\Omega_0$ & $\gamma$ & $Po$ & $Re_0$ & $Re_f$ & $Ek$ & $Ro_f$\\ \hline
\midrule

\REF1{MHD00} & \REF1{8} & \REF1{3} & \REF1{$3.75\times10^{-1}$} & \REF1{$1.3\times10^{3}$} & \REF1{$6.6\times10^{2}$} & \REF1{$1.25\times10^{-4}$} & \REF1{$8.2\times10^{-2}$} \\
MHD01 & 10 & 0.1 & $1.0\times10^{-2}$ & $1.3\times10^{3}$ & $2.9\times10^{2}$  & $1.0\times10^{-4}$ & $2.9\times10^{-2}$ \\
MHD02 & 10 & 1 & $1.0\times10^{-1}$& $1.3\times10^{3}$ & $3.7\times10^{2}$ & $1.0\times10^{-4}$ & $3.7\times10^{-2}$ \\
MHD03 & 10 & 10 & $1.0\times10^{0}$ & $1.3\times10^{3}$ & $1.24\times10^{3}$ & $1.0\times10^{-4}$ & $1.24\times10^{-1}$ \\
MHD04 & 1 & -1 & $1.0\times10^{0}$ & $1.3\times10^{3}$ & $6.3\times10^{2}$ & $1.0\times10^{-3}$ & $6.3\times10^{-1}$ \\
MHD05 & 1 & -3 & $3.0\times10^{0}$ & $1.3\times10^{3}$ & $4.2\times10^{2}$ & $1.0\times10^{-3}$  & $4.2\times10^{-1}$ \\
MHD06 & 1 & -5 & $5.0\times10^{0}$ & $1.3\times10^{3}$ & $3.85\times10^{2}$ & $1.0\times10^{-3}$  & $3.85\times10^{-1}$ \\
MHD07 & 8 & -1 & $1.25\times10^{-1}$ & $1.3\times10^{3}$ & $4.1\times10^{2}$ & $1.25\times10^{-4}$ & $5.1\times10^{-2}$\\
MHD08 & 8 & -3 & $3.75\times10^{-1}$ & $1.3\times10^{3}$ & $1.3\times10^{3}$ & $1.25\times10^{-4}$ & $1.6\times10^{-1}$ \\
MHD09 & 8 & -3.5 & $4.4\times10^{-1}$ & $1.3\times10^{3}$ & $1.6\times10^{3}$ & $1.25\times10^{-4}$ & $2.0\times10^{-1}$ \\
MHD10 & 8 & -4 & $5.0\times10^{-1}$  & $1.3\times10^{3}$ & $1.9\times10^{3}$ & $1.25\times10^{-4}$ & $2.4\times10^{-1}$\\
MHD11 & 8 & -4.5 & $5.6\times10^{-1}$ & $1.3\times10^{3}$ & $2.3\times10^{3}$ & $1.25\times10^{-4}$ & $2.9\times10^{-1}$\\
MHD12 & 8 & -5 & $6.25\times10^{-1}$ & $1.3\times10^{3}$ & $2.8\times10^{3}$ & $1.25\times10^{-4}$ & $3.5\times10^{-1}$\\

MHD13 & 16 & -1 & $6.2\times10^{-2}$ & $1.3\times10^{3}$ & $4.2\times10^{2}$ & $6.25\times10^{-5}$ & $2.6\times10^{-2}$\\
MHD14 & 16 & -3 & $1.9\times10^{-1}$ & $1.3\times10^{3}$ &$1.1\times10^{3}$ & $6.25\times10^{-5}$ & $6.85\times10^{-2}$\\
MHD15 & 16 & -3.5 & $2.2\times10^{-1}$ & $1.3\times10^{3}$ & $1.3\times10^{3}$ & $6.25\times10^{-5}$ & $8.35\times10^{-2}$\\
MHD16 & 16 & -4 & $2.5\times10^{-1}$ & $1.3\times10^{3}$ & $1.6\times10^{3}$ & $6.25\times10^{-5}$ & $1.0\times10^{-1}$\\
MHD17 & 16 & -4.5 & $2.8\times10^{-1}$ & $1.3\times10^{3}$ & $1.9\times10^{3}$ & $6.25\times10^{-5}$ & $1.2\times10^{-1}$\\
MHD18 & 16 & -5 & $3.1\times10^{-1}$ & $1..3\times10^{3}$ & $2.2\times10^{3}$ & $6.25\times10^{-5}$ & $1.4\times10^{-1}$\\
\bottomrule
\bottomrule
\end{tabular*}
\caption{Information of the MHD simulations with their name ID and adimensional parameters. The viscosity is the same for all runs and, in consequence, the initial Reynolds number too.}
\label{table_MHD}
\end{center}
\end{table}

\section{\label{sec:Results}Results}

\subsection{\label{subsec:Results_HD}HD Results}

We first report the results for the purely hydrodynamic case (i.e., no magnetic field). 
We performed a parametric study changing three of the parameters of the system: the rotation rate $\Omega_0$, the precession frequency $\gamma$ and the kinematic viscous coefficient $\nu$.  

\begin{figure}[H]
    \centering
    \includegraphics[scale=0.9]{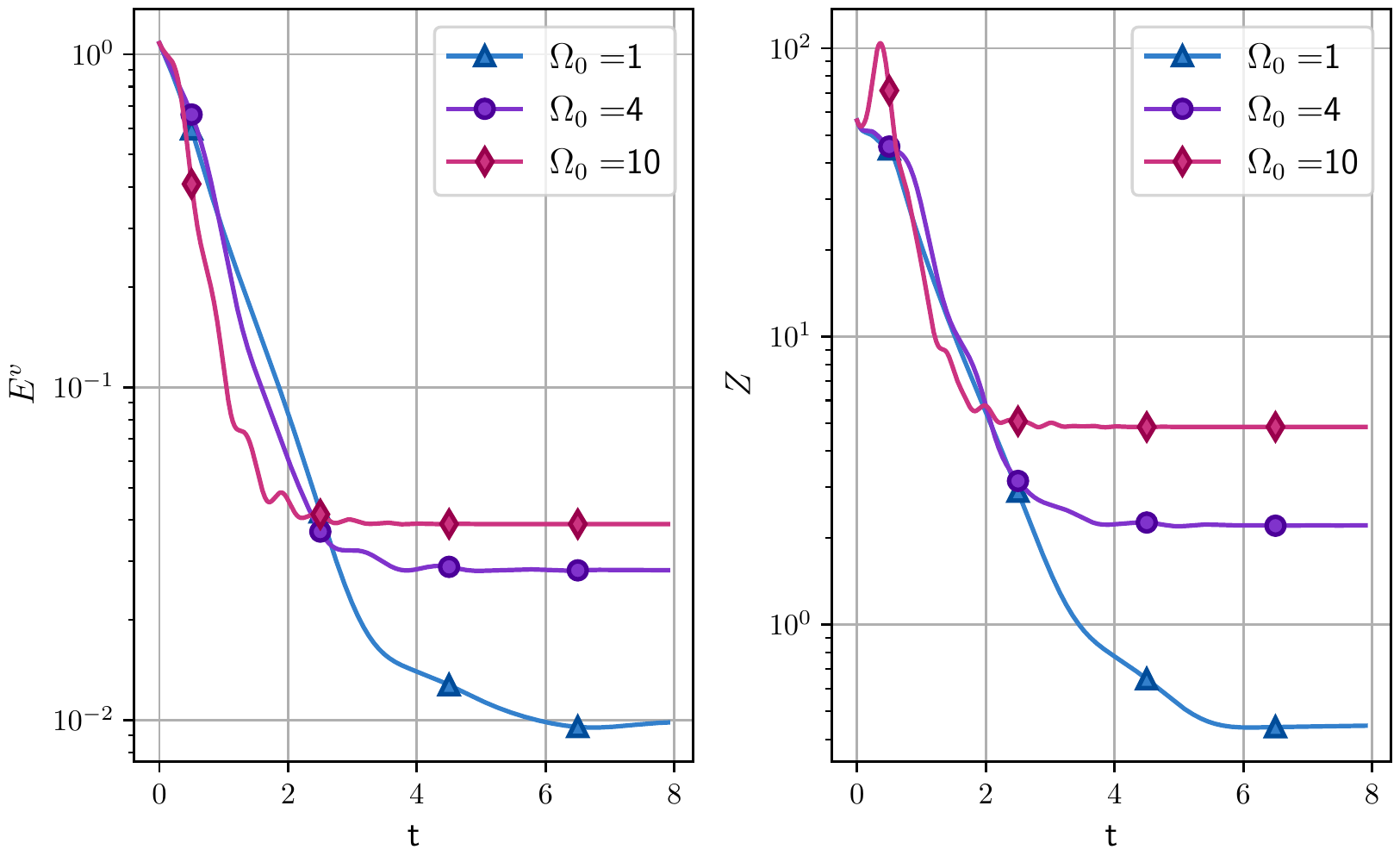}
    \caption{Energy $E_k$ (left) and averaged enstrophy $\langle \omega^2 \rangle$ (right) as a function of time for the runs HD04, HD05 and HD06.}
    \label{ene_enst_HD_prec}
\end{figure}

Since the torque given by the precession is forcing the system, whereas viscosity is acting as a dissipation force, it is natural to expect that at certain time the energy and enstrophy have a statistically stabilization value. This is evidenced in Figure \ref{ene_enst_HD_prec} where for the two quantities a transitory regime followed by a stabilization is observed for our simulations. The transient initial decaying regime observed here is similar to previous results for the case without precession \cite{Mininni2007}. On the other hand, the presence of precession gives an asymptotic value for both the energy $\int_V \boldsymbol u^2 dV$ and enstrophy $\int_V \boldsymbol \omega^2 dV$. This asymptotic value is different for each simulation, increasing when either $\Omega_0$ or $\gamma$ are increased (i.e. more energy injected) and decreasing for larger values of $\nu$ (i.e. more energy dissipated).

In an attempt to model this asymptotic scaling we argue the following: looking at the energy equation \ref{energy_balance} without the current density that appears in the MHD case, we consider a scaling for the dissipation rate 
$D\sim \nu U^2/l^2$ where we chose $U=\langle v \rangle$ as a characteristic velocity of the system, and with the Taylor scale as the characteristic length $l\sim\sqrt{\nu R/U}$. This last expression is the result of considering the sphere radius ($R=1$) as the injection length scale in the expression $l \sim Re^{-1/2} R$.
In the scaling for the dissipation rate we neglect any effect of the boundary term.
On the other hand we take a scaling for the energy injection rate as coming uniquely from the precession term,
$\epsilon \sim R U \gamma \Omega_0$. 
Assuming a balance between the dissipation and the injection rate in the
statistically stationary state $D\sim\epsilon$
it follows then that the energy scales as  $U^2\sim\gamma\Omega_0$.

\begin{figure}[H]
    \centering
    \includegraphics[scale=0.9]{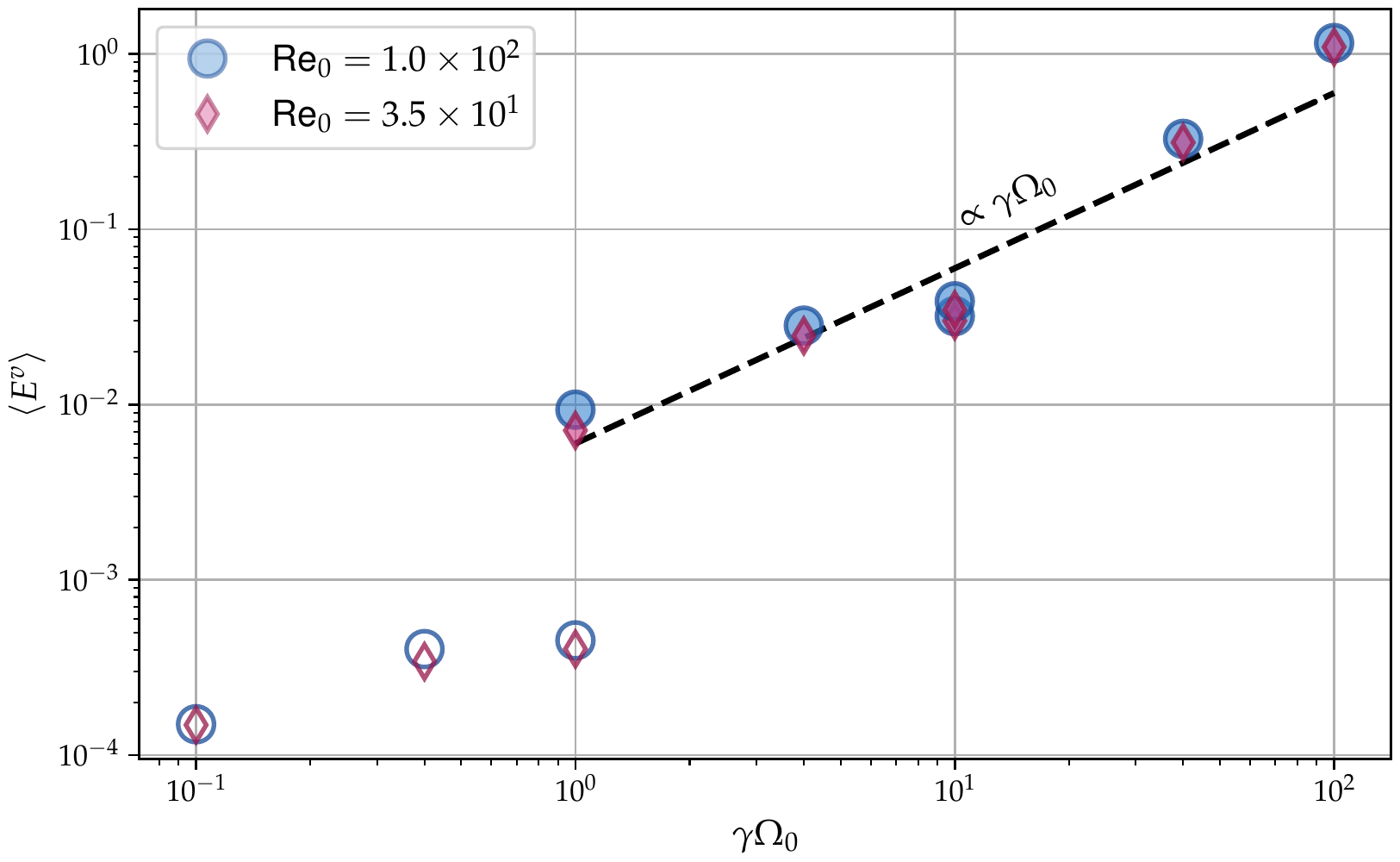}
    \caption{Mean kinetic energy $\langle E^v \rangle$ of all the runs in Table \ref{table_HD} as a function of $\gamma \Omega_0$. The dashed line corresponds to the scaling of the proposed model and filed and open symbols represent the runs with $\gamma \ge 1$ and $\gamma=0.1$ respectively.}
    \label{HD_scaling}
\end{figure}

The proposed scaling is evaluated in Figure \ref{HD_scaling} where we show the mean kinetic energy versus the corresponding $\gamma\Omega_0$ value for that run and compare with the model (represented by a linear behavior in this plot). It can be seen that the results obtained in most of the simulations \REF1{(displayed in the figure with filled symbols)} are consistent with the scaling proposed. \REF1{Furthermore, it can be noticed that the scaling does not match the results for the runs with $\gamma=0.1$, represented in the figure with open symbols. This is expected because the precession frequency is relatively small in comparison with the rotation rate $\Omega_0$.  In this case the hypothesis that the energy injection occurs solely due to the precession term may not be correct in this region of the parameter space.}

\begin{figure}[H]
    \centering
    \includegraphics[scale=0.9]{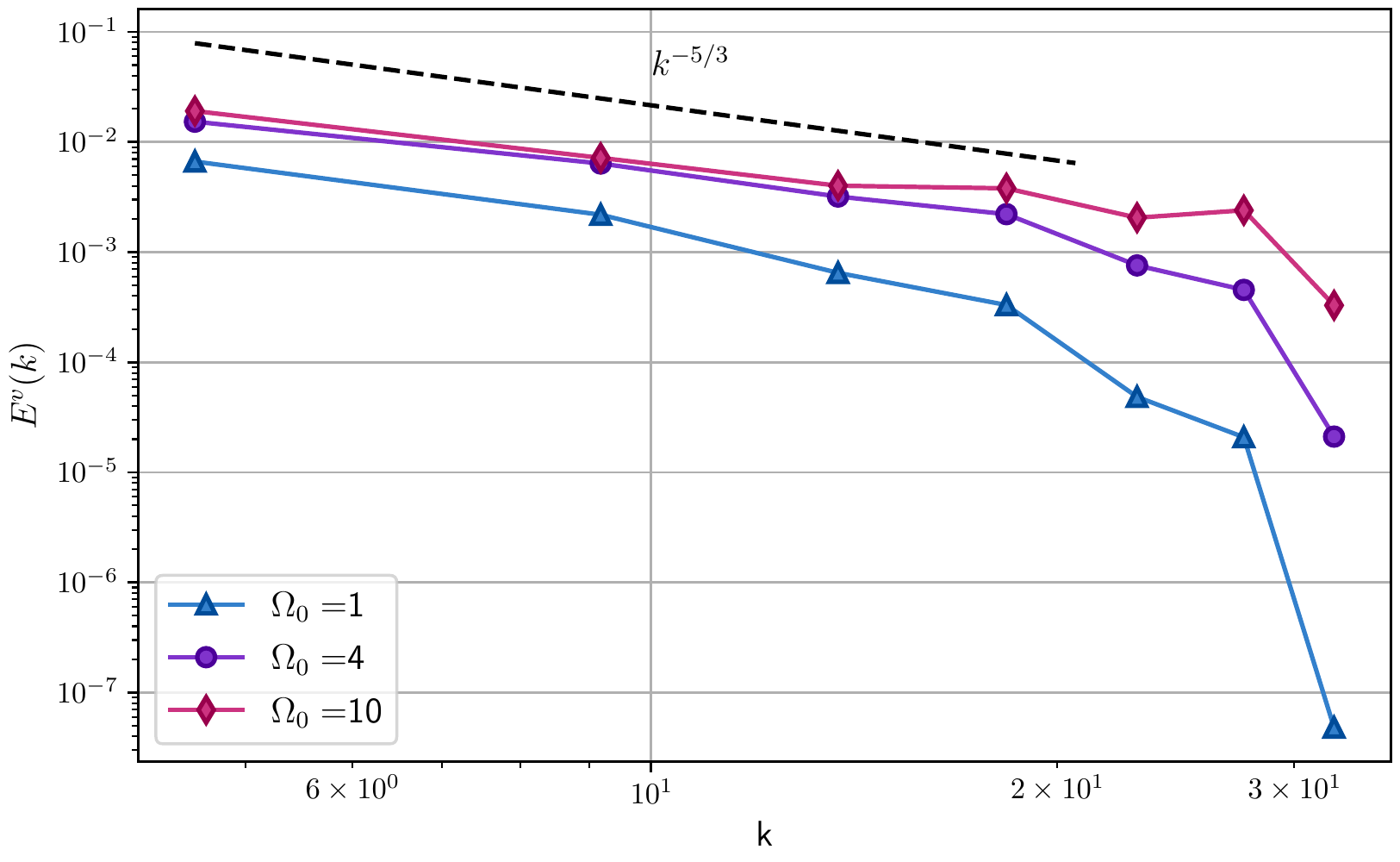}
    \caption{Kinetic power spectra $E$ as a function of the CK wavenumber $k$ for different runs wih varying $\Omega_0$. 
    The reference line corresponds to the Kolmogorov spectrum $k^{-5/3}$.}
    \label{spectra_HD}
\end{figure}

As a final diagnosis for the pure hydrodynamic case in Figure \ref{spectra_HD} we show the kinetic energy spectra for different runs varying $\Omega_0$.  It can be seen that the
spectra follow a Kolmogorov-like behavior ($\sim k^{-5/3}$) at the large scales, which is consistent with the development of stationary turbulence, although the range of scales is very limited.

\REF1{A related subject is the development of (inverse or direct) energy and/or enstrophy cascades. Previous studies have addressed this issue in presence of rotation.
In two-dimensional turbulence it was observed both an inverse cascade for the energy and a direct cascade for the enstrophy \cite{Mininni2013,Sen2012}. For the three-dimensional case a split-cascade in the energy appears whereas the helicity (integral of velocity dot vorticity $\int_V \boldsymbol u \cdot \boldsymbol \omega ~dV$) presents a direct cascade \cite{Teitelbaum2009}. It is also possible the existence of a cascade in the angular momentum when the boundary is closed \cite{Lopez2013}.
As far as we are aware of there is no study of the cascades in presence of precession. This would be an interesting subject to address in future works. }

\subsection{\label{subsec:Results_MHD}MHD Results}

We now consider adding a initial magnetic field to act as a seed, in order to study the feasibility of dynamo action, as well as the existence
of magnetic dipole reversals. The forcing is again given by the precession term and we vary the parameters, in this case $\Omega_0$ and 
the precession frequency $\gamma$ in order to study the different behavior of the system for each case. 

\begin{figure}[H]
    \centering
    \includegraphics[scale=0.9]{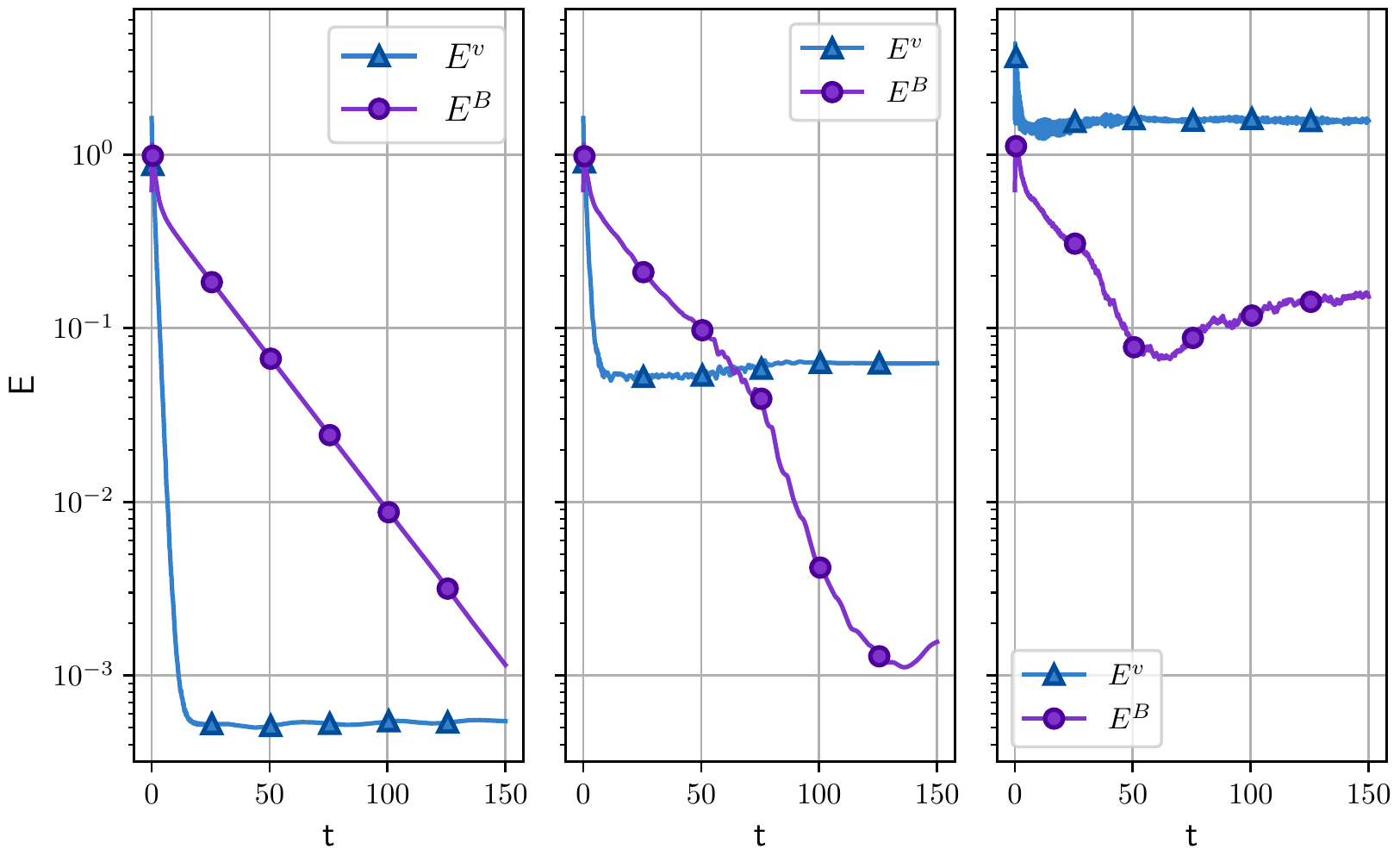}
    \caption{Magnetic and kinetic energy as a function of time for MHD01 (left), MHD02 (middle) and MHD03 (right) which differ in the precession frequency, $\gamma=0.1, 1, 10$, respectively, and have the same value of $\Omega_0=10$. It can be observed that these runs do not generate dynamos, except for very high values of $\Omega_0$ and $\gamma$ as is the case for
    the MHD03 run (rightmost panel).}
    \label{energies_MHD_prograde}
\end{figure}

In Figure \ref{energies_MHD_prograde} we show the results for the kinetic and magnetic energy as a function of time for three different runs, labeled MHD01, MHD02 and MHD03 which differ in the
value of the precession frequency $\gamma=0.1, 1, 10$, respectively, and with the same value for the
rotating frequency $\Omega_0=10$. As it can be observed, the magnetic energy is not sustained, decaying to very low values, except for the high $\gamma$ case, where nevertheless a final 
stationary value lower than the initial value is reached. 

\begin{figure}[H]
    \centering
    \includegraphics[scale=0.9]{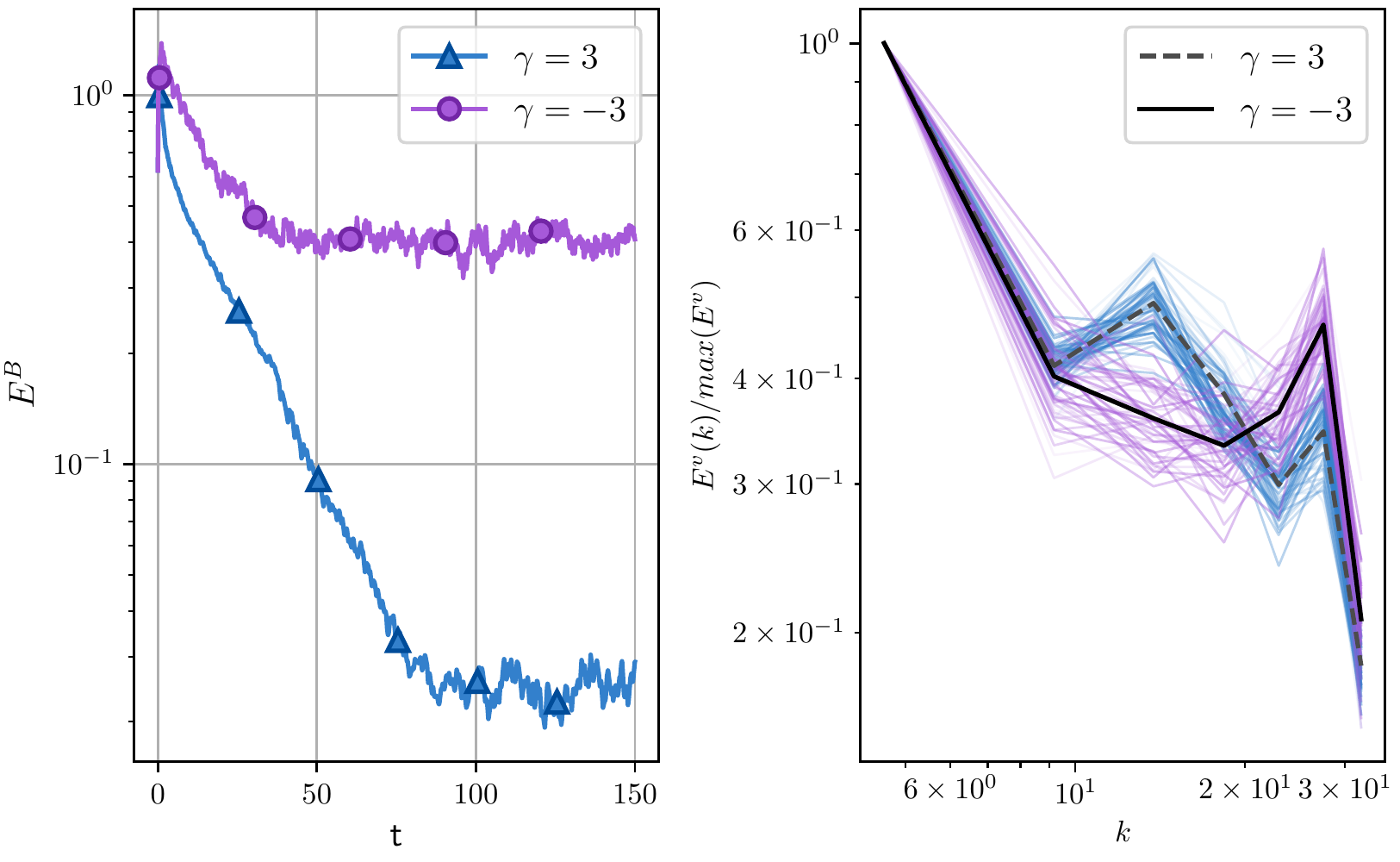}
    \caption{\REF1{Magnetic energy vs time for the prograde and retrogade cases (left); normalized kinetic energy spectra for both cases (right). The spectrum at different times are shown in colors and the time average is shown in dashed line for the prograde case and in solid line for the retrogade case.}}
    \label{MHD_comparation}
\end{figure}

\REF1{We then considered negative values of $\gamma$, that is retrogade precession. A comparison of the magnetic energy vs time for both cases (prograde and retrogade precession) is shown in the left panel of Fig. \ref{MHD_comparation}. It can be seen that in the retrograde case the magnetic energy reaches a higher statistically stationary value than in the prograde case and it also attains it faster. The right panel of the Figure \ref{MHD_comparation} shows the normalized kinetic energy spectra for both cases. The temporal average of the prograde case is in dashed line and the retrograde one is in solid line. It is noted a dominance of the smaller scales in the retrograde case as
compared with the prograde case. This indicates that a stronger turbulent regime is reached in
the retrogade case and this favors the development of small-scale dynamos, as it will be more clearly shown later.}

\begin{figure}[H]
    \centering
    \includegraphics[scale=0.9]{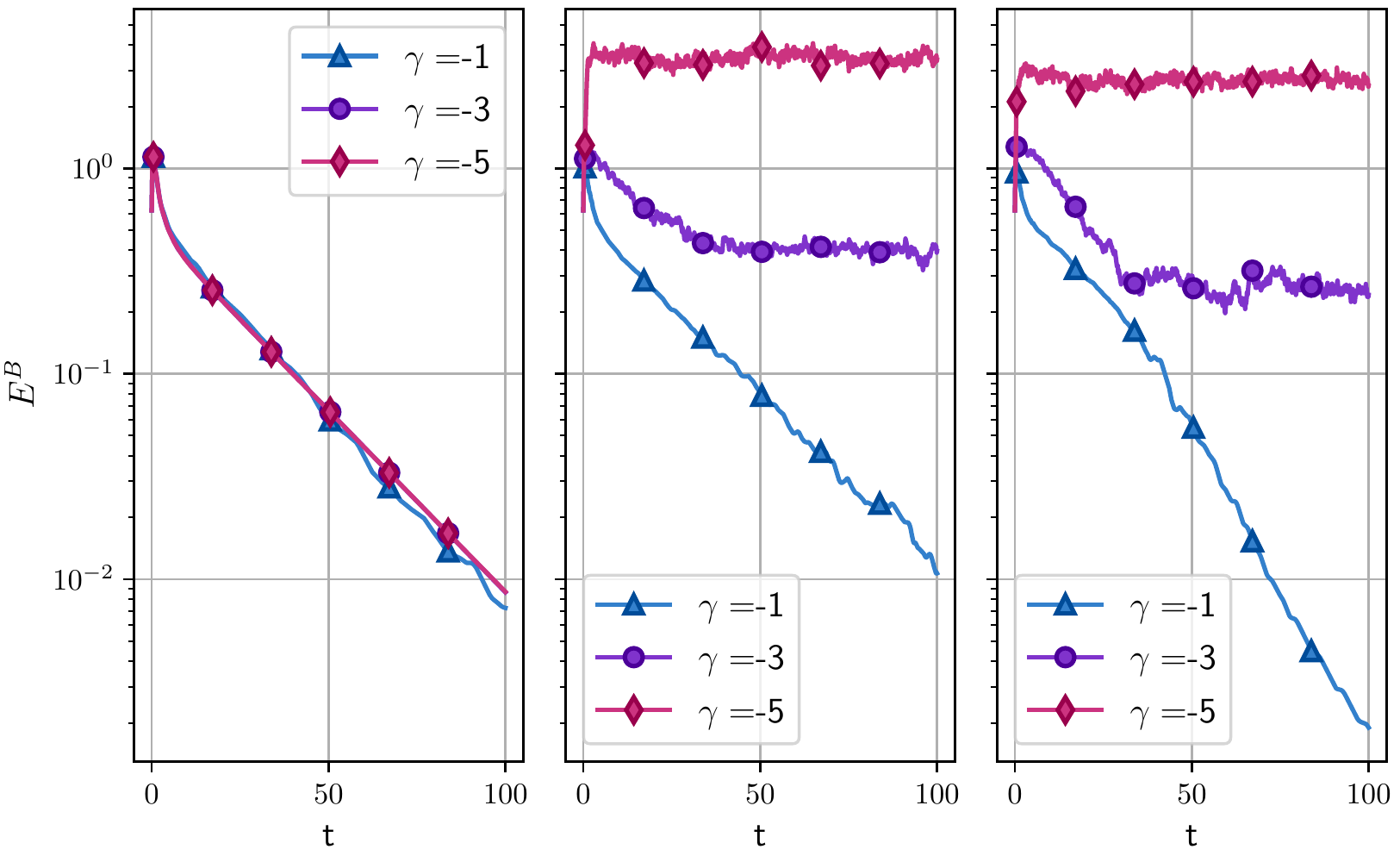}
    \caption{Comparison of the time evolution of the magnetic energy for different $\gamma$ values. Each panel is distinguished by their corresponding value of $\Omega_0$. On the left with $\Omega_0=1$ are the simulations MHD04, MHD05, MHD06, on the middle with $\Omega_0=8$ are MHD07, MHD08 and MHD12, and finaly on the right with $\Omega_0=16$ are the runs MHD13, MHD14 and MHD18.}
    \label{energy_retrograde_MHD}
\end{figure}

\REF1{From now on the work will be focused in the retrograde precession.} This case
is particularly interesting, as it has been 
experimentally studied in several works related to the Earth's liquid core \cite{Vanyo1995,Pais2001,Noir2003,stewartson1963}. 
The results for different runs with retrograde precession are shown in Fig. \ref{energy_retrograde_MHD}. 
The leftmost panel, where the magnetic energy is not sustained, corresponds to the case with $\Omega_0=1$.
On the other hand, in the middle panel with $\Omega_0=8$
dynamo action is observed, for the larger values (in absolute value) of $\gamma=-3, -5$.
A similar result is obtained for the case with $\Omega_0=16$ in the rightmost panel. 
A case of clear magnetic field generation (i.e. dynamo action) is observed for the $\Omega_0 > 1$ cases and for the
largest values of (negative) $\gamma$. \REF1{A critical precession frequency $\gamma_c$ can be also appreciated, which separates the self-sustaining from extinguishing dynamo regimes. This sharp transition can be found in the range $-3<\gamma_c<-1$. Only for $|\gamma|>$  $|\gamma_c|$ is the magnetic energy maintained at a stable level, i.e, a self-sustaining dynamo is attained.}

\begin{figure}[H]
    \centering
    \includegraphics[scale=0.9]{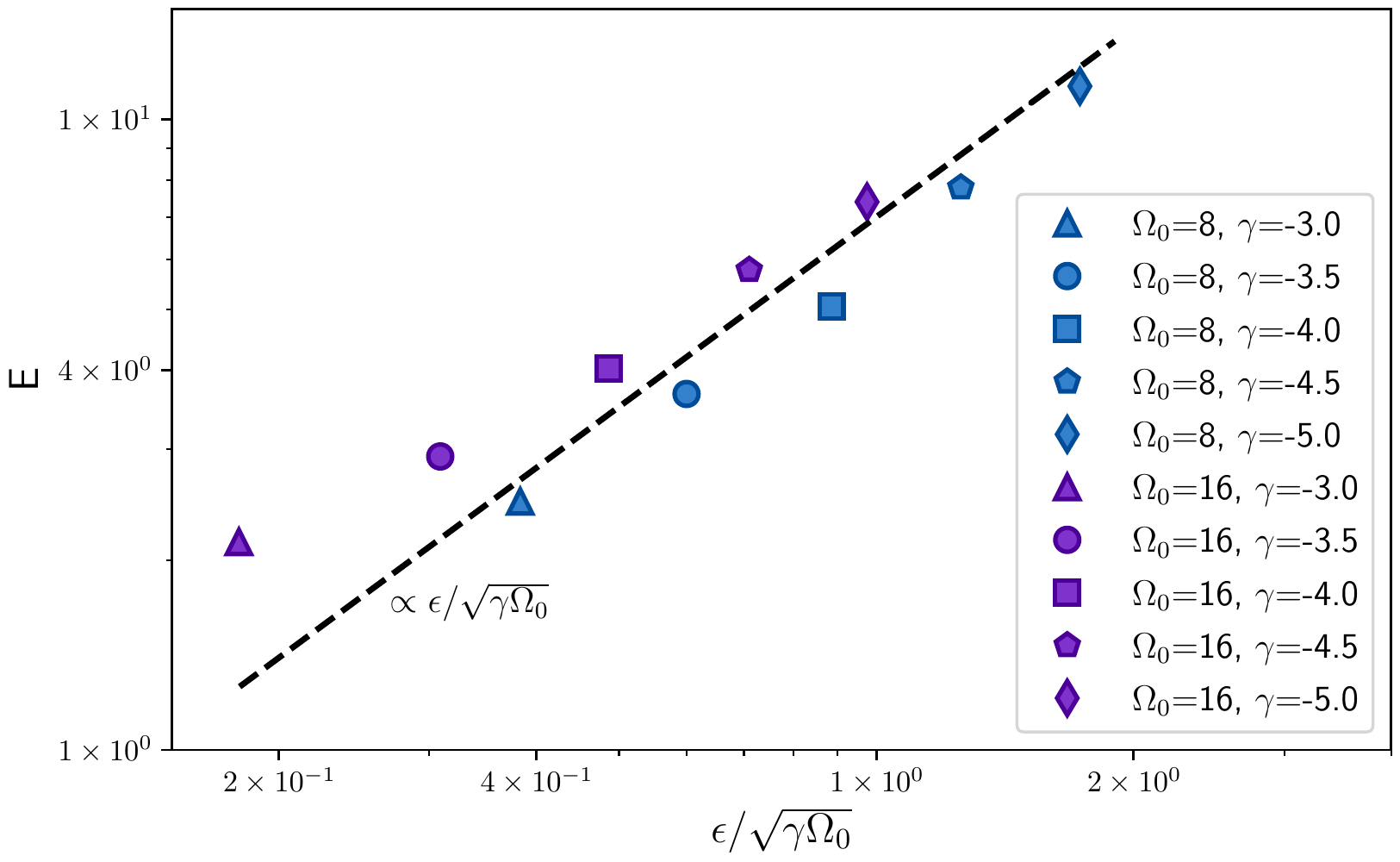}
    \caption{Total energy as a function of the quantity $\epsilon/\sqrt{\gamma \Omega_0}$ for different runs (MHD08-MHD12 and MHD14-MHD18). The proposed scaling is indicated with the dashed line.}
    \label{scaling_energy_MHD}
\end{figure}

In order to obtain a scaling for the total energy (kinetic plus magnetic) we proceed in 
a similar way as was presented in the hydrodynamic results subsection. 
We assume in this case the injection rate is the same as the energy transfer rate (in a statistically stationary state), so $\epsilon \sim (U^3+B^2U)/l \sim U^3/l$, taking
$U\sim B$ as the results of the runs seems to indicate. Here $l$ is a characteristic
length which can be linked with the injection rate and the rotation parameters from the scaling $\epsilon \sim l \Omega_0 \gamma$ so
we obtain $l \sim \epsilon / (\Omega_0 \gamma )$. 
Replacing this scaling for $l$ in the expression
$\epsilon \sim U^3/l$ 
it follows then that the total energy scales as 
$E = U^2+B^2 \sim 2 U^2 \sim \epsilon / \sqrt{\gamma \Omega_0}$.
This scaling is analyzed in Fig. \ref{scaling_energy_MHD}, where the final average energy $E$ is plotted against $\epsilon/ \sqrt{\gamma \Omega_0}$ for several runs. The straight line corresponding to the perfect scaling is shown as a reference and it is notable that for this case a turbulent scaling reproduces satisfactorily the behavior. 

\begin{figure}[H]
    \centering
    \includegraphics[scale=0.9]{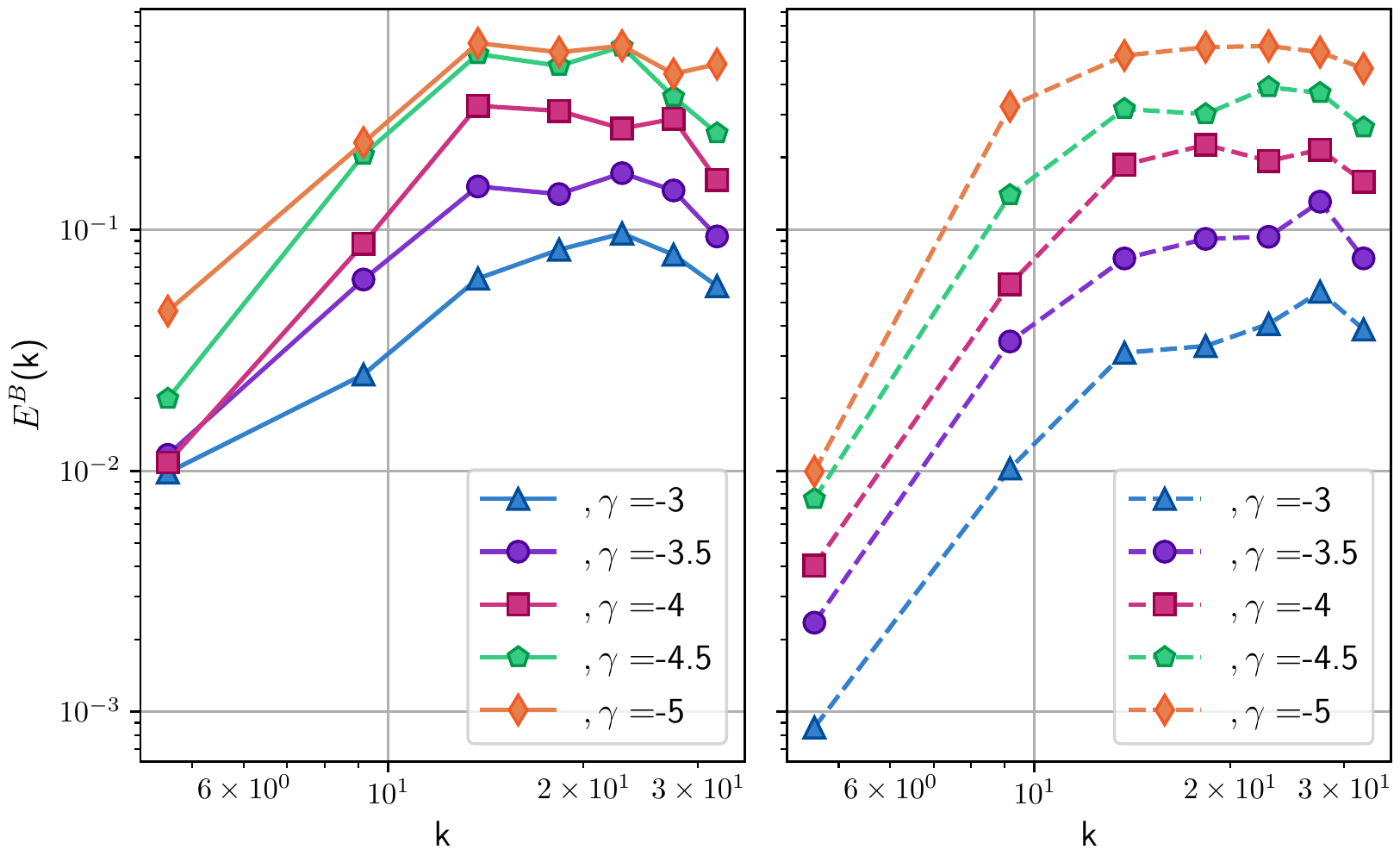}
    \caption{Magnetic energy spatial spectra for runs with $\Omega_0=8$ (MHD08-MHD12) on the left and with $\Omega_0=16$ (MHD14-MHD18) on the right, all of them are for the time $t=150$. It can be observed a predominance in the high values of $k$ for all of the runs. This suggests that these are small-scale dynamos.}
    \label{MHD_spectra}
\end{figure}

Another type of diagnosis to characterize the topology of the magnetic field is presented in Fig. \ref{MHD_spectra}, which shows the
magnetic energy spectra for different runs, all with statistically stationary magnetic energy (i.e. self-sustaining dynamos). 
As can be seen the magnetic energy seems to be dominated by the larger values of $k$,
meaning the dynamos are in a small-scale regime. 

\begin{figure}[H]
    \centering
    \includegraphics[scale=0.9]{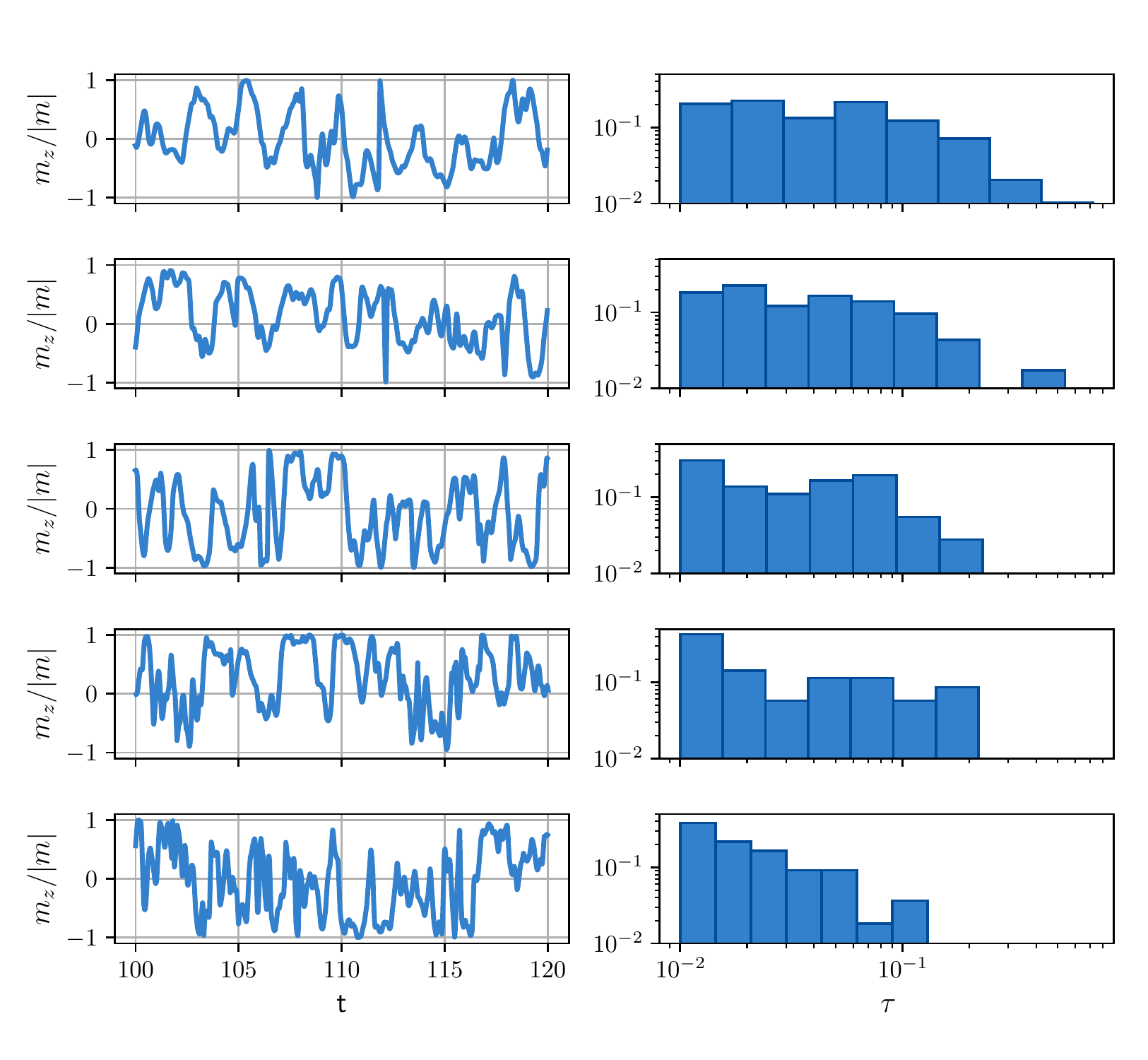}
    \caption{Time evolution of the normalized $z$ component of the magnetic dipole moment 
    $m_z/|m|$ for a time window between t=100 and t=120, for all the dynamo runs with $\Omega_0=8$ (left) and values of  $\gamma=-3, -3.5, -4, -4.5, -5$, from top to bottom. Normalized histograms of the time $\tau$ between reversals (waiting times) are shown on the right. The simulations that are shown in this figure are MHD08-MHD12.}
    \label{MHD_reversals}
\end{figure}

An interesting phenomenon are the reversals of the magnetic dipole moment component parallel to the mean rotation axis, $m_z$. We show the behavior of the normalized value $m_z/|m|$ vs time in Figure \ref{MHD_reversals} for a short time 
window (to better appreciate the dynamics) for a set of runs with $\Omega_0=8$
and different values of $\gamma=-3, -3.5, -4, -4.5, -5$, from top to bottom.
The plots reveal the existence of reversals in all these dynamo runs and it can also be seen
that the dynamics seems to be faster for the larger values of $|\gamma|$. The right panel 
of Fig.  \ref{MHD_reversals} shows  the normalized histograms corresponding to the distribution of times between reversals ({\it waiting times}). The histograms are consistent with
the fact that the reversals occurs faster when the value of the precession frequency $|\gamma|$ is greater, because there is a greater domain of long waiting times for cases with smaller values of $|\gamma|$. 

\begin{figure}[H]
    \centering
    \includegraphics[scale=0.9]{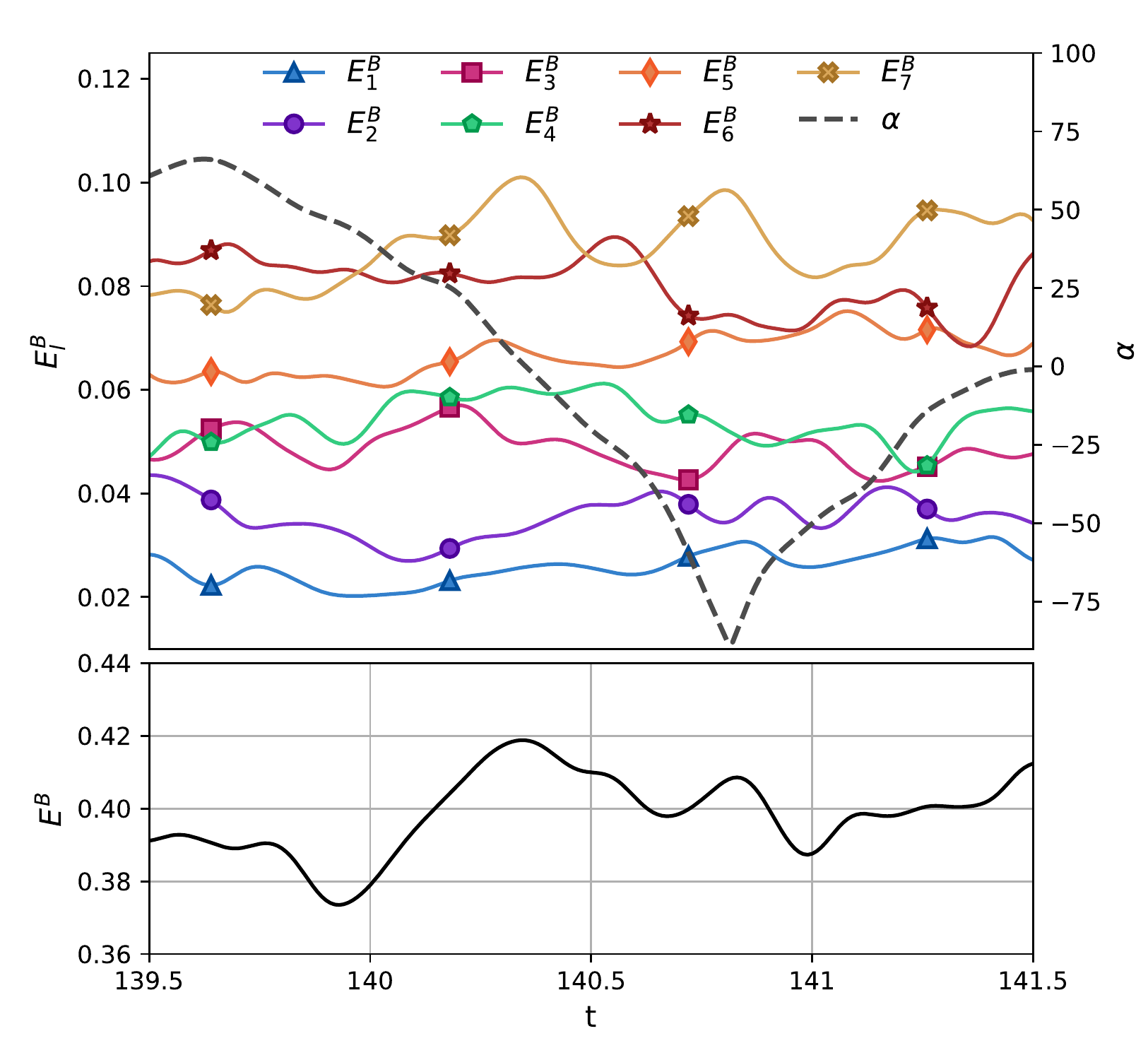}
    \caption{\REF1{Top panel: Magnetic energy evolution during a reversal, at time $t\approx 140.4$, of the different orders of the magnetic multipole expansion ($E_l^B$) labelled with the subscript $l$, in different symbols/colors. The dashed line indicates the evolution of the magnetic dipole latitude ($\alpha$). Bottom panel: total magnetic energy ($E^B$) as a function of time during the same interval. This plot corresponds to the run MHD08.}}
    \label{MHD_magnetic_moments}
\end{figure}

\REF1{We analyzed the dynamics of the magnetic field during a reversal. For this purpose we estimated the amount of magnetic energy for each spherical harmonic degree $E^B_l$ as a function of time. The result for a reversal during run MHD08 is shown in the top panel of Fig.~ \ref{MHD_magnetic_moments} together with the magnetic dipole latitude $\alpha$. It can be readily observed that the magnetic energy in the higher harmonics is greater than in the lower orders, a feature which is consistent with a small-scale dynamo scenario discussed regarding Fig.~ \ref{MHD_spectra}. Furthermore this organization of the magnetic energy seems to remain even when a reversal occurs. This behavior was consistently found in others reversals for this run as well as other operation parameters within the self-sustaining dynamo regime. In the bottom panel of Fig.~ \ref{MHD_magnetic_moments}, the total magnetic energy $E^B$ during the reversal is shown. It can be observed that $E^B$ maintains an approximately steady value during the whole interval, a finding which is consistent with previous studies in other dynamos regimes (see e.g. \cite{Fontana2018} for the case of large scale dynamos). It can be therefore concluded that the dynamo structure does not seem to be sensitive to the reversal of the dipolar moment.} 

\section{Discussion}\label{conclusions}

In the scenario of a rotating and precessive sphere filled with fluid, the incompressible HD and MHD equations were studied by direct numerical simulations. For both cases, a Galerkin spectral code was used. This method decomposes the fields in orthogonal Chandrasekhar-Kendall eigenfunctions. Being purely spectral, this numerical technique allowed us to integrate the system with high accuracy.

In a purely hidrodynamic setting, we found that the kinetic energy and enstrophy present a transitory regime that behaves like the non-precessional case, after which a steady state is reached. This regime was dominated by dissipation even though the spectrum present a power law of $k^{-5/3}$ which corresponds to the development of turbulence. We presented a model for the scaling of the kinetic energy with 
the parameters of the system which showed very good agreement with the results from the simulations. 

For the MHD case, we separated the study into in two scenarios, by taking into account the direction of precession rotation. For the case of prograde precession we could generate dynamos for high values of $\Omega_0$ and $\gamma$. \REF1{On the other hand
using retrograde precession (negative sign of the precession frequency) generates a solution with more intense turbulence in the smallest scales and this favors the self-sustaining dynamos.} For this case of retrograde precession if the $\Omega_0$ value is large ($\Omega_0 = 8$ or $\Omega_0 = 16$) we showed that there is a critical value of the precession frequency in order to obtain a stabilization in the magnetic energy. This critical frequency is \REF1{$-3<\gamma_c<-1$} for either of the two values considered for the angular velocity. Flows operating at $\gamma < \gamma_c$ showcased self-sustaining dynamos in both cases. For low rotation amplitude ($\Omega_0=1$) a monotonic decay of the magnetic energy was observed in the region of parameter space explored.
For the MHD cases which presented sustained dynamo action, a scaling relation for the total energy in the steady state was proposed using the parameters of the problem.
The proposed scaling showed a very good agreement with the results as well. 

Finally we studied the behaviour of the generated magnetic field. We found that all dynamos present dominance of the small-scales in the spectra of the magnetic energy suggesting that the simulations operate in the regime of small-scale dynamos. Another interesting feature of all dynamos with a preferential direction is the presence of magnetic dipole moment reversals. Moreover, doing a statistical analysis, we observed slower dynamics in the dipole moment when the precession is also slower, i.e., when the absolute value of the frequency precession ($|\gamma|$) is smaller. \REF1{Furthermore, we showed that the contribution of each spherical harmonic degree $l$ to the magnetic energy remains equally structured over time, in a statistical sense, as previously found for small-scale dynamos \cite{Fontana2018}.}

\REF1{The presented results constitutes a new contribution to the study of the influence of the precession in 
the turbulent dynamics of HD and MHD in a rotating sphere filled with a fluid or magnetofluid. We believe that these results show interesting and novel features of rotating fluids with precession, including the ability to generate self-sustaning MHD dynamos with magnetic dipole reversals.}

\begin{acknowledgments}

The authors acknowledge support from CONICET and ANPCyT through PIP, Argentina Grant No. 11220150100324CO, and PICT, Argentina Grant No. 2018-4298.

\end{acknowledgments}
\bibliography{references}
\end{document}